\PassOptionsToPackage{dvipsnames}{xcolor}
\documentclass[aps,prb,notitlepage,twocolumn]{revtex4-2}
\usepackage{amsmath}
\usepackage{amsfonts}
\usepackage{amssymb}
\usepackage{subdepth}
\usepackage{graphicx}
\usepackage{slashed}
\usepackage{bm}
\graphicspath{{Plots/}}
\usepackage{mystyleold} 
\usepackage{accents} 
\renewcommand{\leq}{\leqslant}
\renewcommand{\geq}{\geqslant}
\newcommand{\bs}{\begin{subequations}}
\newcommand{\es}{\end{subequations}}
\newcommand{\adb}{\allowdisplaybreaks } 
\newcommand{\ann}{\adb \nonumber \\}
\newcommand{\tm}{\mathrm{tm}}
\newcommand{\te}{\mathrm{te}}
\newcommand{\gr}{\mathrm{gr}}
\newcommand{\x}{\mathsf{x}}
\newcommand{\ii}{\mathrm{i}} 
\newcommand{\n}{\mathsf{n}} 
\newcommand{\dd}{\mathrm{d}}
\graphicspath{{Plots/}}
\DeclareMathOperator{\tr}{tr}
\DeclareMathOperator{\diag}{diag}

\newcommand{\Vl}[1]{\stackrel{_\leftarrow}{#1}}
\newcommand{\Vr}[1]{\stackrel{_\rightarrow}{#1}}

\newcommand{\Eb}{\mathbf{E}}
\newcommand{\Mb}{\mathbf{M}}
\newcommand{\Tb}{\mathbf{T}}
\newcommand{\Rb}{\mathbf{R}}
\newcommand{\Kb}{\mathbf{K}}

\newcommand{\Ab}{\mathbf{A}}
\newcommand{\Bb}{\mathbf{B}}
\newcommand{\Cb}{\mathbf{C}}

\newcommand{\Gb}{\mathbf{G}}
\newcommand{\Wb}{\mathbf{W}}
\newcommand{\Sb}{\bm{\mathcal{S}}}
\newcommand{\Ec }{\mathcal{E}}
\newcommand{\Fc }{\mathcal{F}}
\newcommand{\Dc}{\bm{\mathcal{D}}}

\newcommand{\rb}{\bm{r}}
\newcommand{\kb}{\bm{k}}
\newcommand{\tb}{\bm{t}}
\newcommand{\bb}{\bm{b}}

\newcommand{\ub}{\bm{u}}
\newcommand{\vb}{\bm{v}}

\newcommand{\alphab}{\bm{\alpha}}
\newcommand{\sigmab}{\bm{\sigma}}
\newcommand{\etab}{\bm{\eta}}
\newcommand{\taub}{\bm{\tau}}
\newcommand{\rhob}{\bm{\rho}}
\newcommand{\phib}{\bm{\phi}}
\newcommand{\psib}{\bm{\psi}}
\newcommand{\Phib}{\bm{\Phi}}
\newcommand{\Psib}{\bm{\Psi}}
\newcommand{\Ib}{\mathbf{I}}
\newcommand{\ov}[1]{\accentset{*}{#1}}
\usepackage{xcolor} \definecolor{darkgreen}{rgb}{0,.5,0} \definecolor{myyel}{HTML}{ffff80}
\usepackage[colorlinks,filecolor=blue,citecolor=darkgreen,unicode]{hyperref}
\begin{document}
\title{The Casimir effect for stack of graphenes}
\author{Natalia Emelianova}	\email{natalia7emelianova@gmail.com}
\affiliation{CMCC, Universidade Federal do ABC, Avenida dos Estados 5001, CEP 09210-580, SP, Brazil}
\author{Rashid Kashapov}\email{rashid.kashapov@gmail.com}
\affiliation{Eidos-Robotics LLC , Peterburgskaya 50, 420107, Kazan, Russia}
\author{Nail Khusnutdinov}\email{nail.khusnutdinov@gmail.com}
\affiliation{CMCC, Universidade Federal do ABC, Avenida dos Estados 5001, CEP 09210-580, SP, Brazil}
	\date{\today}
	\begin{abstract}
		We consider a  stack of parallel sheets composed of conducting planes with tensorial conductivities. Using the scattering matrix approach, we derive explicit formulas for the Casimir energy of two, three, and four planes, as well as a recurrence relation for arbitrary planes. Specifically, for a stack of graphene, we solve the recurrence relations and obtain formulas for the Casimir energy and force acting on the planes within the stack. Moreover, we calculate the binding energy in the graphene stack with graphite interplane separation, which amounts to $E_{ib} = 9.9$ meV/atom. Notably, the Casimir force on graphene sheets decreases rapidly for planes beyond the first one. In particular, for the second graphene layer in the stack, the force is $35$ times smaller than that experienced by the first layer.			
	\end{abstract}  
\maketitle 
	
\section{Introduction} 

The Casimir effect \cite{Casimir:1948:otabtpcp}, which was originally considered for two perfect slabs, now plays an important role in various phenomena in physics, chemistry, and biology. For instance, books \cite{Bordag:2009:ACE,*Milton:2001:CEPMZE,*Parsegian:2006:VdWFHBCEP,*Milonni:1994:QVItQE} and recent reviews \cite{Woods:2016:MpoCavdWi} have highlighted its significance in these fields. The Casimir force, which can have different values and signs depending on the types of materials used, such as graphene \cite{Bordag:2009:CibapcagdbtDm}, topological insulators \cite{Grushin:2011:tcrwtdti,*Lu:2018:cdWtfbatis}, chiral metamaterials \cite{Zhao:2009:rCfcm}, Weyl semimetals \cite{Wilson:2015:rcfbws,*RodriguezLopez:2020:socoriciotiaiws}, the shape of material boundaries, and the external conditions including temperature, chemical potential, and magnetic field, has demonstrated its importance in chemistry \cite{Sheehan:2009:cc}, biophysics \cite{Machta:2012:ccfcm,*Pawlowski:2013:qcemufobscm}, and in layered systems such as graphite.

Since the 1970s, van der Waals/Casimir energy and force have been studied for multilayered periodic systems, including a periodic stack of dielectric materials \cite{Ninham:1970:vdWIiMS,*Ninham:1970:vdWFaTF}. The Lifshitz formula has been generalized for layered dielectric systems using nonstandard recursion relations for Fresnel coefficients and the fluctuation-dissipative theorem \cite{Tomas:2002:cfam,*Tomas:2012:cealm}. The Casimir force acting on a plane in a stack perpendicular to the $z$-axis was calculated as the difference in the $(z, z)$ component of the regularized stress tensor of neighbouring planes. In Ref.\,\cite{Teo:2010:cprmiamm}, a plane in a stack was considered as a piston for a five-plane system. The Casimir force for a five-layered magnetodielectric planar system was discussed in Ref.\,\cite{Ellingsen:2007:cfrmscg} as an ideal system for detection of the temperature dependence of the Casimir force, and a path-integral approach was used in Ref.\,\cite{Kheirandish:2011:cfpmlms,*Amooghorban:2011:cfmmbgl} to calculate the force for a magnetodielectric layered planar system. Different geometries of multilayered systems were considered in Ref.\,\cite{Sernelius:2014:EnmaCeils} in the framework of normal-mode techniques, and a modal approach was developed for layered materials in Ref.\,\cite{Davids:2010:macfps}. Finally, Ref.\,\cite{Abbas:2017:staemotcfigm} explored the thermal and electrostatic manipulation of the Casimir force in graphene-dielectric multilayers and demonstrated the possibility of consistent modulation of the Casimir pressure by changing the number of graphene sheets in the stack. 

The study presented herein deals with the calculation of Casimir energy for a layered system with a different shape, i.e., a set of cylindrical concentric shells, which has been previously discussed in Ref.\,\cite{Tatur:2008:ZeNpcccs}. In Ref.\,\cite{Khusnutdinov:2015:cefacopcs,*Khusnutdinov:2015:cefasocp,*Khusnutdinov:2016:cpefasocp,*Kashapov:2016:tcefpls,*Khusnutdinov:2018:tcaciinp2dm,*Khusnutdinov:2019:cei2dmss}, the layered system of conductive planes was analysed in details. Two models of conductivity were taken into consideration, namely the constant conductivity and the Drude--Lorentz 7-oscillator model of graphene conductivity, based on the conductivity of graphite \cite{Djurisic:1999:Opog}.

It is important to note that, in all of the aforementioned references, the scalar type of materials was considered. Conversely, the conductivity of graphene has a tensorial form \cite{Bordag:2009:CibapcagdbtDm,Fialkovsky:2011:FCefg,*Bordag:2016:ECefdg}. For two planes with matrix Fresnel reflection coefficients, the formula for the Casimir energy, including some variations and a correct derivation, can be found in Ref.\,\cite{Bordag:1995:Veisbf,*Reynaud:2010:sacf,*Ingold:2015:cesa} and Ref.\,\cite{Fialkovsky:2018:qfcrbcss}, respectively. In a system consisting of two planes without virtual photon production, this formula can be represented as an integral over imaginary frequency $\xi = -\ii\omega$, 
\begin{equation}
	E_2 = \Re \iint \frac{\dd^2 k}{(2\pi)^3} \int_0^\infty \dd \xi \ln \det \left[\Ib - e^{-2 d k_E} \rb'_1 \rb_2\right],
\end{equation} 
where $\rb'_1$ and $\rb_2$ are the Fresnel reflection matrices for the first and second planes, $d$ represents the interplane distance, and $k_E = \sqrt{\xi^2 + \kb^2}$. The generalization of this expression for more plains is not obvious and requires careful derivation, which is the main focus of this paper.

In Sec. \ref{Sec:Scatt}, the scattering problem for a layered system containing $\n$ parallel conductive planes with tensorial conductivity is formulated. A general expression of Casimir's energy is obtained, and it is calculated for $\n = 2, 3$, and $4$ planes with different conductivities and interplane distances. 

In Sec. \ref{Sec:Scatt}, recurrent relations for the energy are derived, and a relation for energies for $\n, \n - 1, \n - 2$ planes is obtained. In the case of a stack of graphenes with equal interplane distance, this recurrent relation can be solved in manifest form. By using eigenvalues of the reflection Fresnel matrix, the matrix form of solutions is represented as a sum of scalar forms over eigenvalues. Analogous to Casimir's energy, the Casimir force acting on a specific plane in the stack is determined.

Finally, Sec. \ref{Sec:Graph} is dedicated to a particular case of a stack of graphenes, where the Casimir energy and force are numerically calculated at zero temperature, taking into account the conductivity tensor determined within the framework of 3D QED \cite{Bordag:2009:CibapcagdbtDm}.

Throughout the paper, the units where $\hbar = c =1$ are used.

\section{Scattering problem}\label{Sec:Scatt}

\begin{figure}
	\centering
	\includegraphics[width=0.7\linewidth]{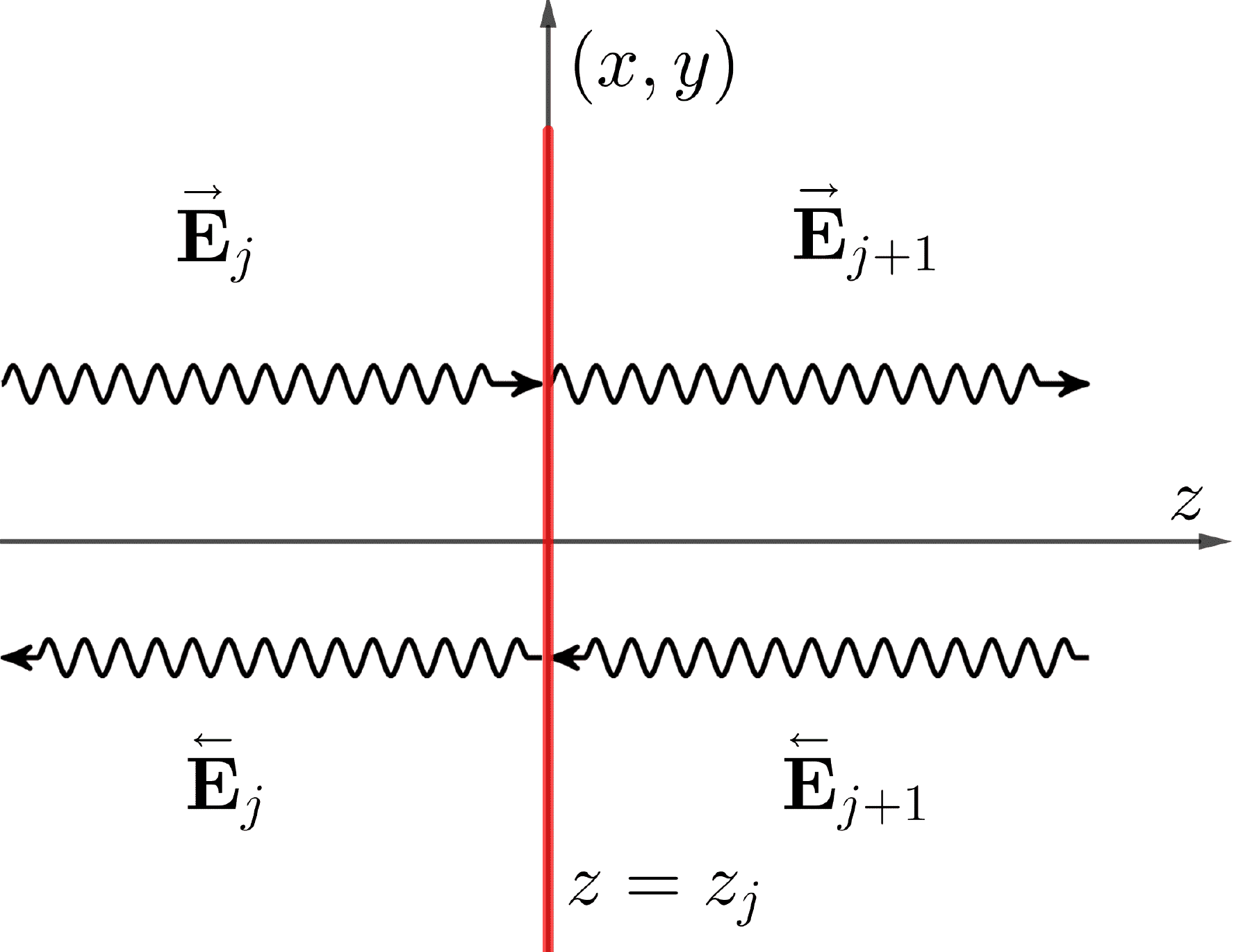}
	\caption{The scattering on the plane $j$ is shown. The waves moving from the left (right) to the right (left) are defined by factor $e^{+\ii k_z z} (e^{-\ii k_z z})$.}
	\label{fig:fields}
\end{figure}

Let us consider the scattering problem for the system of $\n$ conductive planes, which are perpendicular to the axis $z$ in the points $z_1 < z_2 < \cdots < z_\n$. This system divides all space on $\n+1$ domains, $(D_1|z_1|D_2|z_2|\ldots,|z_\n|D_{\n+1})$. For a plane $j$ at position $z=z_j$ (see Fig.\,\ref{fig:fields}) we have the following scattering problem
\begin{equation}\label{eq:Sj}
	\begin{pmatrix}
		\Vl{\Eb}_{j}\\
		\Vr{\Eb}_{j+1}
	\end{pmatrix} =
	   \Sb_j \begin{pmatrix}
		\Vr{\Eb}_{j}\\
		\Vl{\Eb}_{j+1}
	\end{pmatrix},
\end{equation}
where 
\begin{equation}
	\Sb_j = \begin{pmatrix}
		\rb_j & \tb'_j\\
		\tb_j & \rb'_j
	\end{pmatrix},
\end{equation}
is the scattering $4\times 4$ matrix on the plain $j$ ($j=1,\ldots,\n$).  Here, $\rb_j$ and $\tb_j$ are the $2\times 2$ reflection and transmission coefficients of plane $j$, correspondingly.  The $\Vr{\Eb}_{j} (\Vl{\Eb}_{j})$ is 2 component vector $(E_x,E_y)^T$ in the domain $D_j$. The movement to the right $\Vr{\Eb}_{j}$ (left $\Vl{\Eb}_{j}$), is defined by factor $e^{+\ii k_z z} (e^{-\ii k_z z})$, 
To simplify notation, we do not write out the factors with a position of planes. They may be restored by replacements 
\begin{equation}
	\rb_j \to \rb_j e^{2\ii k_z z_j},\ \rb'_j \to \rb'_j e^{-2\ii k_z z_j},\  \tb_j \to \tb_j ,\   \tb'_j \to \tb'_j . \label{eq:replacement}
\end{equation}

The scattering matrix of total system, $\Sb$, is defined by relation 
\begin{equation}\label{eq:scattmatrix}
	\begin{pmatrix}
		\Vl{\Eb}_{1}\\
		\Vr{\Eb}_{\n+1}
	\end{pmatrix} =
	\Sb \begin{pmatrix}
		\Vr{\Eb}_{1}\\
		\Vl{\Eb}_{\n+1}
	\end{pmatrix} = \begin{pmatrix}
	\Rb & \Tb'\\
	\Tb & \Rb'
\end{pmatrix} \begin{pmatrix}
\Vr{\Eb}_{1}\\
\Vl{\Eb}_{\n+1}
\end{pmatrix}.
\end{equation}
To obtain matrix $\Sb$, we represent the set of connected scattering equations \eqref{eq:Sj} into form $\Mb\cdot \Eb = 0$, where 
\begin{equation}
	\Mb =
	\begin{pmatrix}
		\Ab_1 & \Cb_1 & 0  & 0 & \cdots & 0 &0\\
		0   & \Ab_2 &\Cb_2 & 0 & \cdots & 0 & 0\\
		0   & 0   &\Ab_3 & \Cb_3 & \cdots &0 & 0\\
		\vdots   & \vdots   & \vdots & \vdots & \ddots &\ddots &\vdots\\
		0   & 0   &0 & 0 & \cdots &\Ab_{\n+1} &\Cb_{\n+1}
	\end{pmatrix},\ \Eb = 
\begin{pmatrix}
	\Eb_1 \\
	\Eb_2 \\
	\Eb_3 \\
	\vdots \\
	\Eb_{\n+1}
\end{pmatrix}, 
\end{equation} 
and
\begin{equation}
	\Ab_i =
	\begin{pmatrix}
		\rb_i & -\Ib\\
		\tb_i & 0
	\end{pmatrix},
	\Cb_i =
	\begin{pmatrix}
		0 & \tb'_i\\
		-\Ib & \rb'_i
	\end{pmatrix}, 
	\Eb_i =
	\begin{pmatrix}
	\Vr{\Eb}_i\\
	 \Vl{\Eb}_i
   \end{pmatrix}.
\end{equation}
Here, $\Ib$ is the identity matrix $2\times 2$. To obtain a relation between $\Eb_1$ and $\Eb_{\n+1}$, we make elementary transformations of lines in the matrix $\Mb$ to the form where the last line has the first and last elements, only. To do this, we sequentially will make zeros blocks, starting with the third line. 

Let us illustrate these transformations by using an example with three planes.  We multiply the second line of $\Mb$ on the left on $-\Ab_3 \Cb_2^{-1}$ and add it to third line of the matrix:
\begin{equation}
	\Mb =
	\begin{pmatrix}
		\Ab_1 & \Cb_1 & 0  & 0 \\
		0   & \Ab_2 &\Cb_2 & 0 \\
		0   & -\Ab_3 \Cb_2^{-1}\Ab_2   &0 & \Cb_3
	\end{pmatrix}.
\end{equation}
Next, we multiply on the left the first line on $\Ab_3\Cb_2^{-1}\Ab_2\Cb_1^{-1}$ and add it to the third line and then multiply the third line on $\Cb_3^{-1}$:
\begin{equation}
	\Mb =\begin{pmatrix}
		\Ab_1 & \Cb_1 & 0  & 0 \\
		0   & \Ab_2 &\Cb_2 & 0 \\
		\Cb_3^{-1}\Ab_3 \Cb_2^{-1} \Ab_2 \Cb_1^{-1}\Ab_1   & 0   &0 & \Ib
	\end{pmatrix}.
\end{equation}
The last line corresponds to equations
\begin{equation}\label{linEq}
	\Kb_3 
	\begin{pmatrix}
		\Vr{\Eb}_1\\
		\Vl{\Eb}_1
	\end{pmatrix} =
	\begin{pmatrix}
		\Vr{\Eb}_4\\
		\Vl{\Eb}_4
	\end{pmatrix}, 
\end{equation}
where
\begin{equation}
	\Kb_3 = 	(\Cb_3^{-1} \Ab_3) (\Cb_2^{-1} \Ab_2) (\Cb_1^{-1} \Ab_1) = \Bb_3 \Bb_2 \Bb_1,
\end{equation}
and $\Bb_i = \Cb_i^{-1} \Ab_i$ (the general structure of these matrices may be found in Appendix \ref{Sec:app1}). For a system with $\n$ planes we obtain 
 \begin{equation}\label{eq:linEqG}
	(-1)^{\n+1}\Kb_\n 
	\begin{pmatrix}
		\Vr{\Eb}_1\\
		\Vl{\Eb}_1
	\end{pmatrix} =
	\begin{pmatrix}
		\Vr{\Eb}_{\n+1}\\
		\Vl{\Eb}_{\n+1}
	\end{pmatrix},
\end{equation}
where 
\begin{equation}\label{eq:Kdef}
	\Kb_\n = \Bb_{\n} \Bb_{\n-1} \cdots \Bb_1,
\end{equation}
are the matrices $4\times 4$. For $\n=0$, there is no scattering at a whole, and therefore $\Kb_0 = \Ib_4$, the identity matrix $4\times 4$. By using simple algebra, we can represent this system in the form \eqref{eq:scattmatrix} and obtain the scattering matrix of the total system in terms of the matrix $\Kb$ (we omit index $\n$ in $\Kb_\n$ for simplicity)
\begin{align}
	\Rb &= -\Kb_{22}^{-1}\Kb_{21},\ \Rb' = \Kb_{12}\Kb_{22}^{-1} ,\ \Tb' = (-1)^{\n+1}\Kb_{22}^{-1},\ann 
	\Tb &= (-1)^{\n+1}\left(\Kb_{11} - \Kb_{12}\Kb_{22}^{-1}\Kb_{21}\right). \label{eq:Fresnel}
\end{align}
For $\n=0$ we obtain $\Rb = \Rb'=0$ and $\Tb=\Tb'=\Ib$ as should be the case. The determinant of this block-matrix reads
\begin{equation}
	\det \Kb = \det(\Kb_{22}) \det (\Kb_{11} - \Kb_{12}\Kb_{22}^{-1}\Kb_{21}) = \frac{\det\Tb}{\det \Tb'}.
\end{equation}

As noted in Ref. \cite{Fialkovsky:2018:qfcrbcss}, due to symmetry, the scattering in opposite directions has to be taken into account. All quantities in the opposite direction we will mark by asterisk over $\ov{g} = g(-k_z)$. In Ref. \cite{Fialkovsky:2018:qfcrbcss}, it was shown that
\begin{align}
	\ov{\Rb}' &=  (\Rb' - \Tb \Rb^{-1} \Tb')^{-1},\	\ov{\Rb} =  (\Rb - \Tb' \Rb'^{-1} \Tb)^{-1},\ann
	\ov{\Tb}' &=  (\Tb - \Rb' \Tb'^{-1} \Rb)^{-1},\ \ov{\Tb}  =  (\Tb' - \Rb \Tb^{-1} \Rb')^{-1},\label{eq:direct}
\end{align}
where the right hand side is calculated for $+k_z$. Changing $k_z \to - k_z$ we obtain the inverse formulas
\begin{align}
	\Rb' &= (\ov{\Rb}' - \ov{\Tb} \ov{\Rb}^{-1} \ov{\Tb}')^{-1},\ \Rb = (\ov{\Rb} - \ov{\Tb}' \ov{\Rb}'^{-1} \ov{\Tb})^{-1},\ann
	\Tb' &= (\ov{\Tb} - \ov{\Rb}' \ov{\Tb}'^{-1} \ov{\Rb})^{-1},\ \Tb  = (\ov{\Tb}' - \ov{\Rb} \ov{\Tb}^{-1} \ov{\Rb}')^{-1}.\label{eq:inverse}
\end{align}
These relations valid for a scattering matrix of each plane, too. 

The Casimir energy is defined by the logarithm of the determinant of the scattering matrix, and it may be represented in the following form \cite{Fialkovsky:2018:qfcrbcss}
\begin{equation}
	\det \Sb =  \frac{\det \Rb}{\det \ov{\Rb}'} = \frac{\det \Tb'}{\det \ov{\Tb}'} .
\end{equation}
By using relation \eqref{eq:Fresnel} we obtain the following formula (see Appendix \ref{Sec:app1})
\begin{equation}\label{eq:SK}
	\det \Sb =  \frac{\det \Tb'}{\det \ov{\Tb}'} = \frac{\det \ov{\Kb}_{22}}{\det \Kb_{22}}.
\end{equation}
The contribution to the Casimir energy,
\begin{equation}
	\ln \det \Sb = \ln\det \ov{\Kb}_{22} - \ln \det \Kb_{22},
\end{equation}
is the difference of scatterings in opposite directions. 

Without changing the Casimir energy, we can multiply the matrix $\Kb_\n$ on the arbitrary non-degenerate matrix $\Wb$, which has no dependence on the positions of planes. We use this freedom and define the matrix   $\Dc_n = \Kb_{22}\Wb$ in that way that $\Dc_\n = \Ib + \ldots$. Then, 
\begin{equation}
	\det\Sb = \frac{\det \ov{\Dc}_\n}{\det \Dc_\n}, 
\end{equation}
and the Casimir energy reads
\begin{equation}\label{eq:CasGen}
	E_\n = \Re\iint \frac{\dd^2 k}{(2\pi)^3} \int_0^\infty \dd \xi \ln\det \Dc_\n,
\end{equation}
where $\Dc_\n$ is calculated at the imaginary axis $\omega = \ii \xi$. 

The special cases with $\n=1,2,3,4$ planes are considered in Appendix \ref{Sec:app2}. They are
\begin{align}
	\Dc_1 &= \Ib, \ann 
	\Dc_2 &= \Dc_2^{(21)} = \Ib - \rb_2 \rb'_1,\ann
	\Dc_3 &= \Dc_2^{(32)}\tb'^{-1}_2\Dc_2^{(21)}\tb'_2 - \rb_3\tb_2\rb'_1\tb'_2,\ann
	\Dc_4 &=  \Dc_2^{(43)}\tb'^{-1}_3 \Dc_2^{(32)} \tb'^{-1}_2\Dc_2^{(21)}\tb'_2\tb'_3 - \Dc_2^{(43)}\tb'^{-1}_3\rb_3\tb_2\rb'_1\tb'_2\tb'_3  \ann
	&-  \rb_4 \tb_3 \rb'_2 \tb'^{-1}_2 \Dc_2^{(21)}\tb'_2\tb'_3 - \rb_4 \tb_3 \tb_2\rb'_1\tb'_2\tb'_3.\label{eq:EcN}
\end{align}
The positions of planes are restored by replacement \eqref{eq:replacement}. Here $\Dc_2^{(il)} = \Ib - \rb_i \rb'_l$ is the matrix for two planes $i$ and $l$ (numeration of planes starts from one for the first plane and up to $\n$ for the last one). We observe in the above expressions that the matrices $\tb$ appear for internal planes, only. The first terms are additive contributions to the energy; the rest terms describe the non-additivity of the Casimir energy. For example, the first term for $\n=3$ planes gives the following contribution
\begin{equation}
\ln \det (\Dc_2^{(32)}\tb'^{-1}_2\Dc_2^{(21)}\tb'_2) = \ln\det \Dc_2^{(32)} + \ln\det \Dc_2^{(21)},
\end{equation}
which is the sum of the contribution of pairs of neighboring planes -- first with the middle and the middle with the last one. The second term in $\Dc_3$ contains reflection matrices of the first and last planes and transmission matrix of the middle plane. It looks like a contribution due to the interaction of the first and last terms through the middle.    

\section{Recurrent relations} \label{Sec:Recurr}

To obtain a recurrent relation for $\Dc_\n$ we use Eq.\,\eqref{eq:Kdef}.  The matrix $\Kb_\n$ has the following form 
\begin{equation}
	\Kb_{\n+1} = \Bb_{\n+1} \Kb_\n.
\end{equation}
By using this relation for index $\n$, $\n +1$ and the manifest form of matrix $\Bb_\n$ \eqref{eq:Bdif} we obtain the recurrent relation for block $(2,2)$ the matrix $\Kb_\n$
\begin{equation}
	\Kb_{\n +2}^{22} = \phib_\n \Kb_{\n +1}^{22}  + \psib_\n \Kb_{\n}^{22}, 
\end{equation}
where 
\begin{align}
	\phib_\n & = - \tb'^{-1}_{\n+2} \left[\Ib - \rb_{\n+2} \left(\rb'_{\n+1} - \tb_{\n +1} \rb^{-1}_{\n +1} \tb'_{\n +1}\right)\right],\ann
	\psib_\n & = - \tb'^{-1}_{\n +2} \rb_{\n +2} \tb_{\n +1} \rb^{-1}_{\n + 1}. 
\end{align}
We define matrices 
\begin{equation}
	\Dc_\n = (-1)^\n \tb'_\n \Kb_\n \tb'_1 \tb'_2 \cdots \tb'_{\n -1} = \Ib + \ldots, \ \Dc_1 = \Dc_0 = \Ib.
\end{equation}
In terms of these matrices, we obtain the relation  we need
\begin{equation}\label{eq:RecRelGen}
	\Dc_{\n +2} = \Phib_\n \Dc_{\n +1}\tb'_{\n +1}  + \Psib_\n \Dc_\n\tb'_{\n} \tb'_{\n +1}, 
\end{equation}
where
\begin{align}
	\Phib_\n  & = (\Ib - \rb_{\n +2} \rb'_{\n +1}) \tb'^{-1}_{\n +1} + \rb_{\n +2} \tb_{\n +1} \rb^{-1}_{\n +1}, \ann
	\Psib_\n  & = - \rb_{\n +2} \tb_{\n +1} \rb^{-1}_{\n + 1} \tb^{-1}_\n.
\end{align}

The scattering matrix for conductive plane $s$, with conductivity  tensor $\etab_s = 2\pi \sigmab_s$ was found in Ref.\,\cite{Fialkovsky:2018:qfcrbcss}:
\begin{align}
	\rb_s &= e^{2\ii k_z z_s}\alphab_s,\ \rb'_s = e^{-2\ii k_z z_s}\alphab_s,\	\tb_s = \Ib + \alphab_s,\  \tb'_s = \tb_s, \ann
	\ov{\rb}_s &=  - e^{-2\ii k_z z_s} (\Ib+2\alphab_s)^{-1} \alphab_s,\ \ov{\rb}'_s = - e^{2\ii k_z z_s} (\Ib+2\alphab_s)^{-1} \alphab_s,\ann
	\ov{\tb}_s &=  (\Ib+2 \alphab_s )^{-1} (\Ib+\alphab_s) ,	\ov{\tb}'_s  =  \ov{\tb}_s ,
\end{align}
where 
\begin{align}
	\alphab_s &= -\frac{\omega^2 \bm{\eta}_s - \kb \otimes \kb\etab_s + \Ib k_z\omega \det\etab_s}{\omega^2 \tr \etab_s - (\kb\kb\etab_s)  + k_z\omega (1+ \det\etab_s) },\ann
	\	[\alphab_s]^i_j &= -\frac{\omega^2 [\etab_s]^i_j - k^i k^l[\etab_s]_{lj} + \delta^i_j k_z\omega \det\etab_s}{\omega^2 \tr \etab_s - (\kb\kb\etab_s)   + k_z\omega (1+ \det\etab_s) },
\end{align}
and $\kb = (k^1,k^2) = (k_1,k_2)$, $k_z = \sqrt{\omega^2 - \kb^2}$. All of these matrices commutate with each other. 

At imaginary axes, $\omega \to \ii \xi\ (k_z \to \ii k_E)$ we obtain
\begin{equation}
	\alphab_s = -\frac{\xi^2 \etab_s + \kb \otimes \kb\etab_s + \Ib k_E\xi \det\etab_s}{\xi^2 \tr \etab_s + (\kb\kb\etab_s)  + k_E \xi (1+ \det\etab_s) },
\end{equation}
where $k_E = \sqrt{\xi^2 + \kb^2}$. 

For identical planes $\alphab_s = \alphab$ for all $s$, and
\begin{align}\label{eq:DcRec}
	\Dc_{\n+2} & = \left[\Ib + e^{-2 k_E d_{\n+2,\n+1}}(\Ib + 2\alphab) \right]\Dc_{\n +1}\ann
	& - e^{-2 k_E d_{\n+2,\n+1}}(\Ib+\alphab)^2\Dc_{\n},
\end{align}
where $d_{i,j}=z_j - z_i$ is the distance between planes $i$ and $j$.

For identical planes $\alphab_s = \alphab$ and equal interplane distances $d_{i,j} = d$ we obtain at the imaginary axis
\begin{equation}\label{eq:RecRelD}
	\Dc_{\n+2}  = \left[\Ib +e^{-2 k_E d}(\Ib + 2\alphab) \right]\Dc_{\n +1} - e^{-2 k_E d}(\Ib+\alphab)^2\Dc_{\n},
\end{equation}
and 
\begin{align}
	\Dc_2 &= \Ib - e^{-2 k_E d} \alphab^2,\ann
	\Dc_3 &= (\Ib - e^{-2 k_E d} \alphab^2)^2 - e^{-4 k_E d} \alphab^2 (\Ib + \alphab)^2,\ann
	\Dc_4 &= (\Ib - e^{-2 k_E d} \alphab^2)^3 - 2 e^{-4 k_E d}(\Ib - e^{-2 k_E d} \alphab^2)\alphab^2 (\Ib + \alphab)^2\ann
	&- e^{-6 k_E d} \alphab^2(\Ib + \alphab)^4. \label{eq:Dcn}
\end{align}

To solve the recurrent relations \eqref{eq:RecRelD} we use a generation function method extended for matrix-valued coefficients.  Let us consider the matrix-valued recurrent relations 
\begin{equation}\label{eq:RecRel}
	\Dc_{\n+2} = \ub \Dc_{\n+1} + \vb \Dc_{n},
\end{equation}
with commutative matrices $\ub$ and $\vb$, $[\ub,\vb]=0$, and a generation function
\begin{equation}
	\Gb = \sum_{s=0}^\infty \Dc_s z^s.
\end{equation} 
Here,  
\begin{equation}
	\ub =  \Ib +e^{-2 k_E d}(\Ib + 2\alphab) , \vb  =- e^{-2 k_E d}(\Ib+\alphab)^2.\label{eq:phipsi}
\end{equation}
Taking into account the recurrent relation \eqref{eq:RecRel}, we obtain
\begin{equation}
	\Gb = \Dc_0 + (\Dc_1 - \ub \Dc_0) z + \ub \Gb z + \vb \Gb z^2.
\end{equation}
Therefore, 
\begin{equation}
	\Gb = \frac{\Dc_0 + (\Dc_1 - \ub \Dc_0) z}{\Ib - \ub z - \vb z^2}.
\end{equation}
Then, we expand over $z$
\begin{equation}\label{eq:MSer}
	(\Ib - \ub z - \vb z^2)^{-1} = \sum_{s=0}^\infty \Mb_ s z^s,
\end{equation}
where 
\begin{equation}\label{eq:Mk}
\Mb_s = \sum_{l=0}^{\left[\frac{s}{2}\right]} \ub^{s-2l} \vb^l C^{s-l}_{l}.
\end{equation}
Here, $[\ldots]$ is integer part, and $C^s_l$ is the binomial coefficient
\begin{equation}
	C^s_l = \frac{s!}{l! (s-l)!}.
\end{equation}
We set $\Mb_{-1} = 0$, and from Eq. \eqref{eq:Mk} we obtain $\Mb_0 = \Ib, \Mb_1 = \ub$.

Taking into account this expansion, we have relation
\begin{equation}
	\Gb = \sum_{s=0}^\infty z^s\left[\Dc_0 \Mb_s+ (\Dc_1 - \ub \Dc_0) \Mb_{s-1} \right]= \sum_{s=0}^\infty z^s \Dc_s.
\end{equation}
The solution of this relation reads
\begin{equation}
	\Dc_s  =  \Dc_0 \Mb_s + (\Dc_1 - \ub \Dc_0) \Mb_{s-1}.
\end{equation}
For $s=0,1$ these relations become identities. 

Therefore, 
\begin{equation}\label{eq:DcnS}
	\Dc_\n  = \Mb_\n  - \Mb_{\n-1} e^{-2 k_E d}(\Ib + 2\alphab).
\end{equation}
Here, $\Dc_0 = \Dc_1 = \Ib$ and $\Dc_2, \Dc_3$ are given by Eq. \eqref{eq:Dcn}. 

If the matrices $\ub$ and $\vb$ are numbers, $u,v$, then the sum \eqref{eq:Mk} maybe calculated in close form 
\begin{equation}\label{eq:Mnum}
	M_\n =  \frac{ \left(u + \sqrt{u^2 + 4 v}\right)^{\n+1} - \left(u - \sqrt{u^2 + 4 v}\right)^{\n+1}}{2^{\n+1}\sqrt{u^2 + 4 v}}.
\end{equation}
If the $\ub$ and $\vb$ are diagonal matrices 
\begin{equation}
	\ub = \diag (u_1, u_2), \ \vb = \diag (v_1,v_2),
\end{equation}
then the matrix $\Mb_s$ is diagonal, too, with diagonal elements 
\begin{equation}
	M_\n^{(i)} =  \frac{ \left(u_i + \sqrt{u_i^2 + 4 v_i}\right)^{\n+1} - \left(u_i - \sqrt{u_i^2 + 4 v_i}\right)^{\n+1}}{2^{\n+1}\sqrt{u_i^2 + 4 v_i}}.
\end{equation}

The matrix $\alphab$ may be diagonalized with eigenvalues $r_\te,r_\tm$, which correspond to the reflection coefficients of TE and TM modes  \cite{Khusnutdinov:2019:lteotcpfefaaiwacp}. Therefore, the matrices $\ub$ and $\vb$ \eqref{eq:phipsi} are diagonalized, too,   and, the matrices $\Dc_n$ are diagonal with eigenvalues 
\begin{equation}
	D_\n^\x  = M_\n^\x  - e^{-2 k_E d} M_{\n-1}^\x(1 + 2r_\x),
\end{equation}
where 
\begin{align}
	M_\n^\x &=  \frac{ \left(u_\x + \sqrt{u_\x^2 + 4 v_\x}\right)^{\n+1} - \left(u_\x - \sqrt{u_\x^2 + 4 v_\x}\right)^{\n+1}}{2^{\n+1}\sqrt{u_\x^2 + 4 v_\x}},\ann
	u_\x & = 1 + e^{-2 k_E d}(1 + 2r_\x) , v_\x  = - e^{-2 k_E d}(1+r_\x)^2,
\end{align}
and $\x=\te,\tm$. 

The matrix $\alphab$ has the following eigenvalues 
\begin{equation}
	r_\te = -\left(1 + \frac{k_E}{\eta \xi }\right)^{-1},\ r_\tm = -\left(1 + \frac{\xi }{\eta  k_E}\right)^{-1},
\end{equation}
for the constant conductivity case $\etab = 2\pi \sigmab = \eta \Ib$, and 
\begin{equation}\label{eq:FresnelGraphene}
	r_\te = -\left(1 + \frac{k_E }{\eta_\gr \tilde{k} \Phi } \right)^{-1}, \ r_\tm = -\left(1 + \frac{\tilde{k}}{\eta_\gr k_E  \Phi }\right)^{-1}, 
\end{equation} 
for a graphene at zero temperature case, 
\begin{equation}\label{eq:eta1}
	\etab = \eta_\gr \frac{\tilde{k}}{\xi} \left(\Ib - v_F^2 \frac{\kb \otimes \kb}{\tilde{k}^2}\right) \Phi \left(\frac{\tilde{k}}{2m}\right),
\end{equation}
where $\eta_\gr = 2\pi \sigma_\gr = \pi e^2/2 = 0.0114$, $v_F$ is the Fermi velocity, and 
\begin{equation*}
	\Phi(y) = \frac{2 }{\pi y}\left\{ 1 + \frac{y^2 - 1}{y}\arctan y\right\},\ \tilde{k} =  \sqrt{\xi^2 + v_F^2 \kb^2}. 
\end{equation*}

To compare with the constant conductivity case \cite{Khusnutdinov:2015:cefasocp} we define a new variable $t_\x$ by relation
\begin{equation}
	r_\x = -\frac{t_\x}{1+t_\x}.
\end{equation}
For the constant conductivity case, 
\begin{equation}\label{eq:tconst}
	\ t_\te = \frac{\eta \xi}{k_E},\ t_\tm = \frac{\eta k_E}{\xi}, 
\end{equation}
and 
\begin{equation}\label{eq:tgraphene}
	\ t_\te =\frac{\eta_\gr \tilde{k} \Phi }{k_E },\ t_\tm = \frac{\eta_\gr k_E  \Phi }{\tilde{k}},
\end{equation}
for the graphene's sheets.  With this definition ($z=k_E d$)
\begin{equation}
	M_\n^\x = \frac{e^{-\n z}}{(1+t_\x)^\n} \frac{1 - f_\x^{2(\n+1)}}{f_\x^\n(1-f_\x^2)}, 
\end{equation}
where  
\begin{equation}
	f_\x = \sqrt{(\cosh z + t_\x \sinh z)^2 -1} + \cosh z + t_\x \sinh z.
\end{equation}

Taking these relations into account, we obtain 
\begin{equation}
	D_\n^\x = -\frac{e^{-\n z}}{f^{\n -1}_\x (1+t_\x)^\n} \left(e^{-z} \frac{1-f^{2\n }_\x}{1-f^2_\x} (1-t_\x) -  \frac{1-f^{2(\n+1)}_\x}{f_\x(1-f^2_\x)}\right).
\end{equation}
After some algebra, we can transform this expression to the form obtained in Ref.\,\cite{Khusnutdinov:2015:cefasocp} for the case of scalar constant  conductivity case
\begin{equation} \label{eq:Dnx}
	D_\n^\x = -\frac{e^{-(\n-1)z}}{f^{\n -2}_\x (1+t_\x)^\n} \left(e^{-z} \frac{1-f^{2(\n-1) }_\x}{1-f^2_\x} - \frac{1+t_\x}{f_\x} \frac{1-f^{2\n}_\x}{1-f^2_\x}\right).
\end{equation}

Therefore, we obtain that for a stack of $\n$ identical conductive planes with conductivity tensor $\sigmab$ and with identical interplane distance $d$, the Casimir energy has the following form 
\begin{equation}\label{eq:EnGen}
	E_\n = E_\n^{\te} + E_\n^{\tm}= \iint \frac{\dd^2 k}{(2\pi)^3} \int_0^\infty \dd \xi \left(\ln D_\n^{\te} + \ln D_\n^{\tm}\right),
\end{equation}
where function $D_\n$ is given by Eq. \eqref{eq:Dnx}. 

For an infinite number of planes, $\n\to\infty$, the Casimir energy is divergent, but the energy per unit plane, $\overline{E}_n = E_\n/\n$, is finite it is given by the same expression with replacement $D_\n^\x \to D^\x$, where
 \begin{equation} \label{eq:Dinfx}
 	D^\x =  \lim_{\n \to \infty} \sqrt[\n]{D_\n^\x} =\frac{e^{-z} f_\x}{1+t_\x}.
 \end{equation} 

Let us analyze the limits of small and large separations. To make this, we change integrand variables, $k_i \to k_i /d$, and $\xi \to \xi/d$, to make integrand dimensionless. The only place with $d$-dependence is the argument of function $\Phi = \Phi(\tilde{k}/2md)$. 

For small separations, $d\to 0$, we observe that $\Phi \to 1$ and the integrand does not depend on the interplane distance and therefore $E_\n \sim 1/d^3$. In the zero order of the Fermi velocity, $v_F=0$, the relations \eqref{eq:tgraphene} are transformed to the constant conductivity case  \eqref{eq:tconst} and the energy coincides with that obtained in Ref.\,\cite{Khusnutdinov:2015:cefasocp} for the constant conductivity case. The Drude--Lorentz model considered in Ref.\,\cite{Khusnutdinov:2015:cefasocp} gives $1/d^{5/2}$ dependence. 

For large separations, $d\to \infty$, the function $\Phi \approx 8 \tilde{k}/(3\pi md) \to 0$. Expanding integrand in Eq. \eqref{eq:EnGen} over $t_\x \ll 1$ we obtain 
\begin{equation}
	E_\n = -\frac{\eta_\gr^2 (n-1)}{225 d^3 (md)^2} \sim \frac{1}{d^5}.
\end{equation}
The dependence $1/(m^2d^5)$ is observed in Ref. \cite{Bordag:2009:CibapcagdbtDm}. For the case of graphene/perfect metal the energy $\sim 1/(md^4)$ 

\begin{figure}
	\centering
	\includegraphics[width=0.53\textwidth]{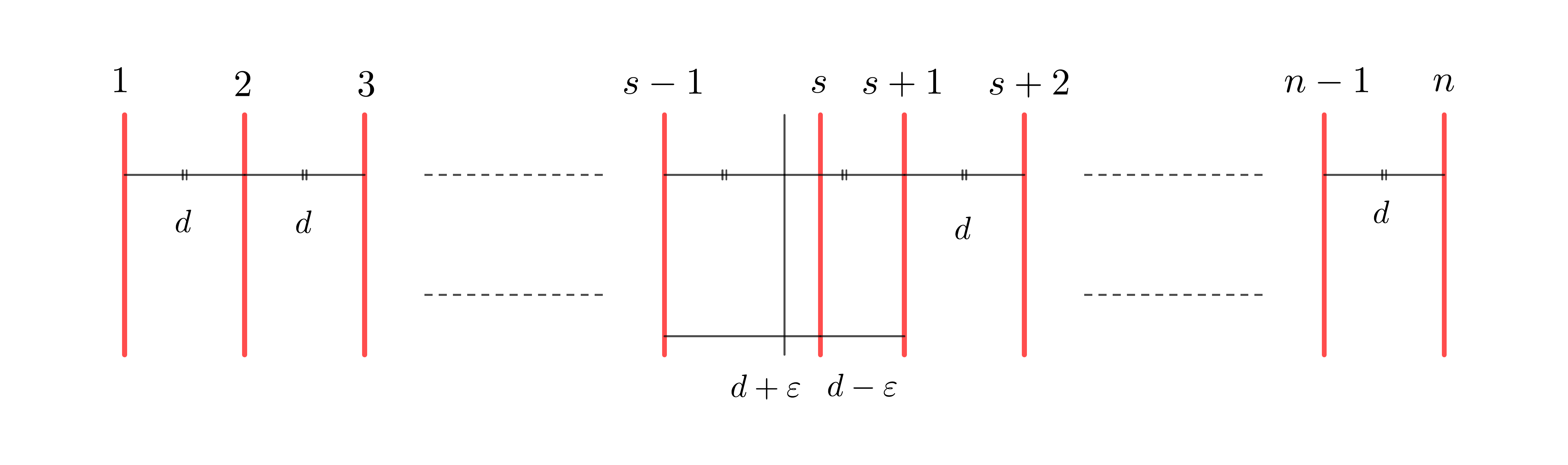}
	\caption{The plane $l=s$ in the stack of $\n$ planes is shifted on $\varepsilon$ to the right.}
	\label{fig:force}
\end{figure}

To obtain a force acting on the plane $s$ in the stack, we consider a stack of graphene with equal distances except for plane $s$, which are shifted on the distance $\varepsilon$ (see Fig.\,\ref{fig:force}) and calculate the derivative
\begin{equation}
	\Dc'_n = \left.\frac{\partial \Dc_n}{\partial \varepsilon} \right|_{\varepsilon =0}.
\end{equation} 
For this system, the recurrent relation \eqref{eq:DcRec} is valid 
\begin{align}
	\Dc_{l}  &=\ub \Dc_{l-1} +\vb\Dc_{l-2}, &1\leq l\leq s-1, \ann
	\Dc_{s}  &=\ub(\varepsilon) \Dc_{s-1} + \vb(\varepsilon) \Dc_{s-2}, &l = s, \ann
	\Dc_{s+1}  &=\ub(-\varepsilon) \Dc_{s} + \vb(-\varepsilon) \Dc_{s-1},& l = s+1, \ann
	\Dc_{l}  &=\ub \Dc_{l-1} +\vb\Dc_{l-2}, &l\geq s+2,
\end{align}
where,
\begin{align*}
	\ub &= \Ib + e^{-2 k_E d}(\Ib + 2\alphab), \vb = - e^{-2 k_E d}(\Ib+\alphab)^2,\ann
	\ub(\varepsilon) &= \left[\Ib + e^{-2 k_E (d+\varepsilon)}(\Ib + 2\alphab) \right], \vb(\varepsilon) = - e^{-2 k_E (d+\varepsilon)}(\Ib+\alphab)^2.
\end{align*}

The recurrent relations for derivatives have the following form
\begin{align}
	\Dc'_{l}  & = 0, &1\leq l\leq s-1, \ann
	\Dc'_{s}  &=\ub' \Dc_{s-1} + \vb' \Dc_{s-2}, &l = s, \ann
	\Dc'_{s+1}  &= -\ub' \Dc_{s} + \ub \Dc_{s} - \vb' \Dc_{s-1} ,& l = s+1, \ann
	\Dc'_{l}  &=\ub \Dc'_{l-1} +\vb\Dc'_{l-2}, &l\geq s+2,
\end{align}
where 
\begin{equation}
	\ub' = -2 k_E e^{-2 k_E d}(\Ib + 2\alphab), \vb' = 2k_E e^{-2 k_E d}(\Ib+\alphab)^2.
\end{equation}

The generation function for this recurrent relation starts from power $z^s$
\begin{equation}
	\Gb = \sum_{l=s}^\infty \Dc'_{l} z^l = z \ub \Gb + z^2 \vb \Gb +  \Dc'_s z^s +( \Dc'_{s+1} - \ub \Dc'_s) z^{s+1}. 
\end{equation}
Therefore,
\begin{equation}
	\Gb = \frac{\Dc'_s z^s +( \Dc'_{s+1} - \ub \Dc'_s) z^{s+1}}{\Ib - z \ub + z^2 \vb}.
\end{equation}
Then we use expansion for denominator \eqref{eq:MSer} and obtain 
\begin{equation}
	\Dc'_\n = \Dc'_s \Mb_{\n-s} + ( \Dc'_{s+1} - \ub \Dc'_s) \Mb_{\n-s-1}.
\end{equation}
For all planes $l \leq s$ we have to use the solution \eqref{eq:DcnS}.

Let us obtain now  an expression for force acting on the graphene with number $s$ in the stack of $\n$ graphenes
\begin{align}
	F_{s,\n} &= - \left.\frac{\partial E_\n}{\partial \varepsilon} \right|_{\varepsilon = 0}  = - \lim_{\varepsilon\to 0} \frac{E_\n(\varepsilon) - E_\n(0)}{\varepsilon} \ann
	&  = - \Re\lim_{\varepsilon\to 0}\iint \frac{\dd^2 k}{(2\pi)^3} \int_0^\infty \dd \xi \frac{\ln\det \Dc_\n(\varepsilon) - \ln\det \Dc_\n(0)}{\varepsilon}\ann
	&=- \Re\iint \frac{\dd^2 k}{(2\pi)^3} \int_0^\infty \dd \xi  \tr \left(\Dc_\n^{-1} \Dc'_\n\right) \ann
	&=- \Re\sum_{\x = \te,\tm}\iint \frac{\dd^2 k}{(2\pi)^3} \int_0^\infty \dd \xi  D^\x_\n{}^{-1} {D^\x_\n}'. \label{eq:Force}
\end{align}
Direct calculation gives 
\begin{equation}\label{eq:FnG}
	D^\x_\n{}^{-1} {D^\x_\n}' \stackrel{\textrm{def}}{=}  G_{s,\n}^\x =  \frac{2z t_\x^2e^{-z} f (f^{2(s-1)} - f^{2(\n -s)})}{e^{-z} f (1-f^{2(\n -1) }_\x) -  (1+t_\x)(1-f^{2\n}_\x)}.
\end{equation}
This expression is the same as obtained in the scalar case in Ref.\,\cite{Kashapov:2016:tcefpls}. For infinite number of graphenes in stack, $\n \to \infty$, and finite $s$ we obtain 
\begin{equation}
	G_s^\x = \frac{2z t_\x^2 f^{2(1-s)}}{1 - e^z f (1+t_\x)}.
\end{equation} 
Therefore, we obtain that the expression for force has the same form as that obtained for the scalar constant conductivity case but with corresponding reflection coefficients for TE and TM modes. 

\section{The stack of graphenes}\label{Sec:Graph}

\begin{figure}
	\centering
	\includegraphics[width=1\linewidth]{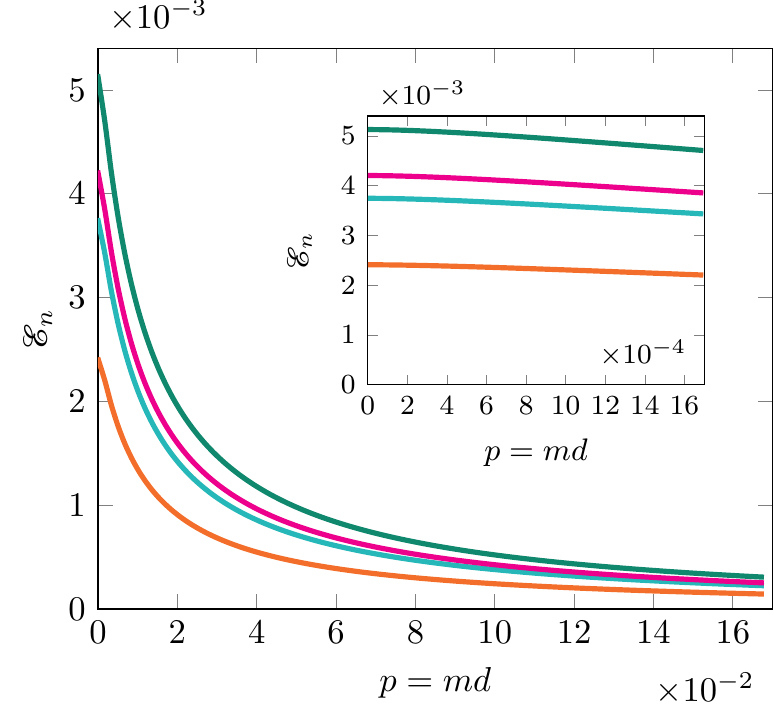}
	\caption{The Casimir energy per unit graphene $\Ec_\n$ as a function of interplane distance. Here, $\n = 2,4,6,\infty$ from the bottom upwards. The insert shows the figure's neighborhood of origin.}
	\label{fig:enp}
\end{figure}

\begin{figure}
	\centering
	\includegraphics[width=1\linewidth]{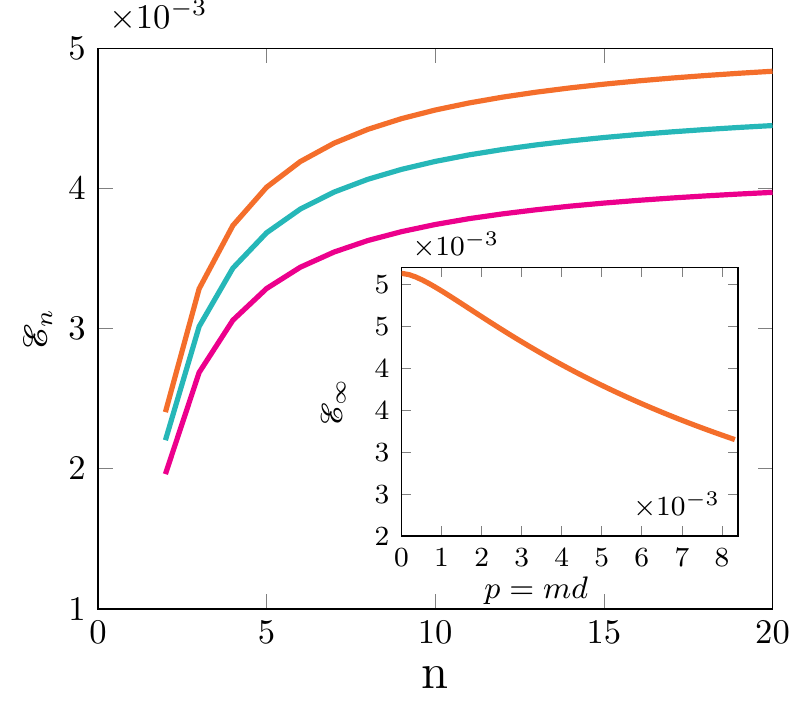}
	\caption{The Casimir energy per unit graphene $\Ec_\n $ as a function of the number of planes in the stack. Here, $m=0.1$ eV, $d = d_c,10 d_c,20 d_c$ from the top down, where $d_c = 0.3345$ is interplane distance in graphite. The insert shows the energy for a stack of an infinite number of graphene as a function of interplane distance up to $50d_c$.}
	\label{fig:enn}
\end{figure}

The numerical evaluation of the Casimir energy for a stack of $\n$ graphenes at zero temperature is plotted in Figs.\,\ref{fig:enp}, \ref{fig:enn} as a function of parameter $p=md$ and number of graphene in the stack.  We evaluate the function $\Ec_\n = E_\n/\n E_{Cas}$ -- the energy per unit graphene in units of the Casimir energy $E_{Cas} = - \pi^2/720 d^3$ for two ideal planes.   

For $m=0.1$ eV and interplane distance in graphite $d_c = 0.3345$ nm, the parameter $p_c=1.69\cdot 10^{-4}$. With this value of mass gap, the domain of distances in Fig.\,\ref{fig:enp} is $d\in [0,100 d_c]$. The energy $\Ec_\n$ is lowered by $16$ times in this interval. For small separation, we observe the $1/d^3$ dependence for the Casimir energy $E_\n$ as was proved above. For large separation, we observe $1/d^5$ dependence of the energy $E_\n$.

The $\n$-dependence of $\Ec_\n$ is plotted in Fig.\,\ref{fig:enn} for different interplane distances.  The energy decreases in accuracy with increasing interplane distance, as should be the case (the insert in Fig.\,\ref{fig:enn}). 

It is useful to compare these calculations with the case of the constant conductivity case with universal conductivity of graphene $\eta_\gr$,  which was considered in detail in Ref.\,\cite{Khusnutdinov:2015:cefasocp}. The constant conductivity leads to $1/d^3$ dependence and therefore $\Ec_\n$ does not depend on the interplane distance. The better agreement appears for small distances, where for the graphene case the $\Ec_\n$ is approximately constant (see Fig.\,\ref{fig:enp}). The case of constant conductivity gives a greater value $\sim 10\%$. 

Let us estimate the binding energy due to the Casimir energy for graphite. For graphite interplane separation $d_c = 0.3345$ nm, we obtain the following energy per unit plane in an infinite stack of graphene $\overline{E}_\infty = 59.23\ \textrm{erg}/\textrm{cm}^2$. The binding energy $E_{ib} = \overline{E}_\infty/d_c \rho_c$, where $\rho_c = 2.23\ \textrm{g}/\textrm{cm}^3$ is the graphite mass density. For these values, we obtain $E_{ib} = 9.9 $ meV/atom, which is by $10$\% smaller than for constant conductivity case \cite{Khusnutdinov:2015:cefasocp}. From the first principle, the cohesion energies are $24-26$ meV/atom \cite{Schabel:1992:Eoibig} and $24$ meV/atom \cite{Rydberg:2003:VdWDFLS}. The experimental data gives cohesion $35\pm 10,\ 15$ meV/atom  \cite{Benedict:1998:Mdotibeig} and $61\pm 5$ meV/atom \cite{Zacharia:2004:Iceogftdoph}. Most likely, the Casimir energy gives an essential contribution to the binding energy. 

\begin{figure}
	\centering
	\includegraphics[width=1\linewidth]{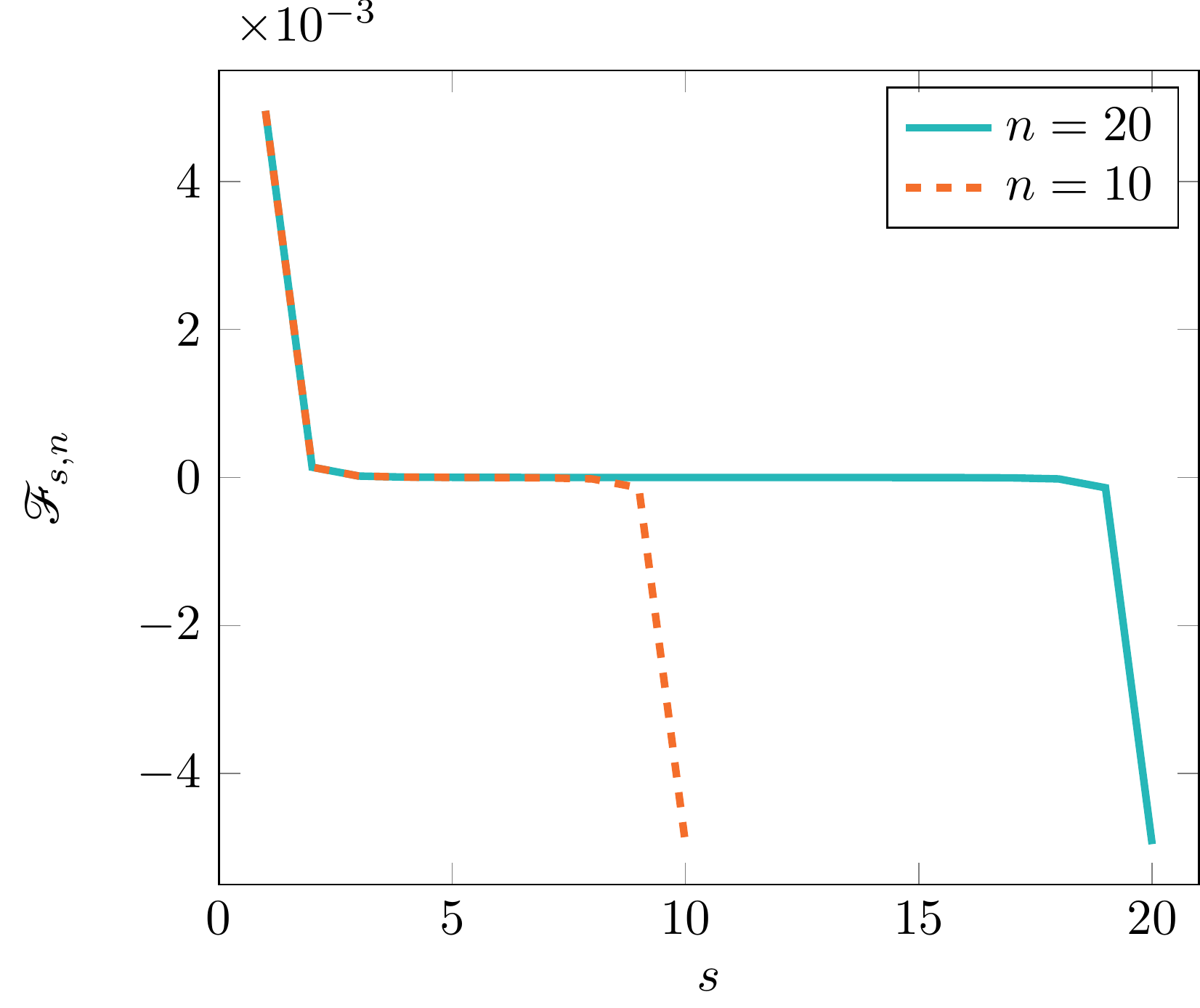}
	\caption{The Casimir force $\Fc_{s,\n}$ acting on the graphene $s$ in the stack of $10$ and $20$ planes. }
	\label{fig:fo1}
\end{figure}

\begin{figure}
	\centering
	\includegraphics[width=1\linewidth]{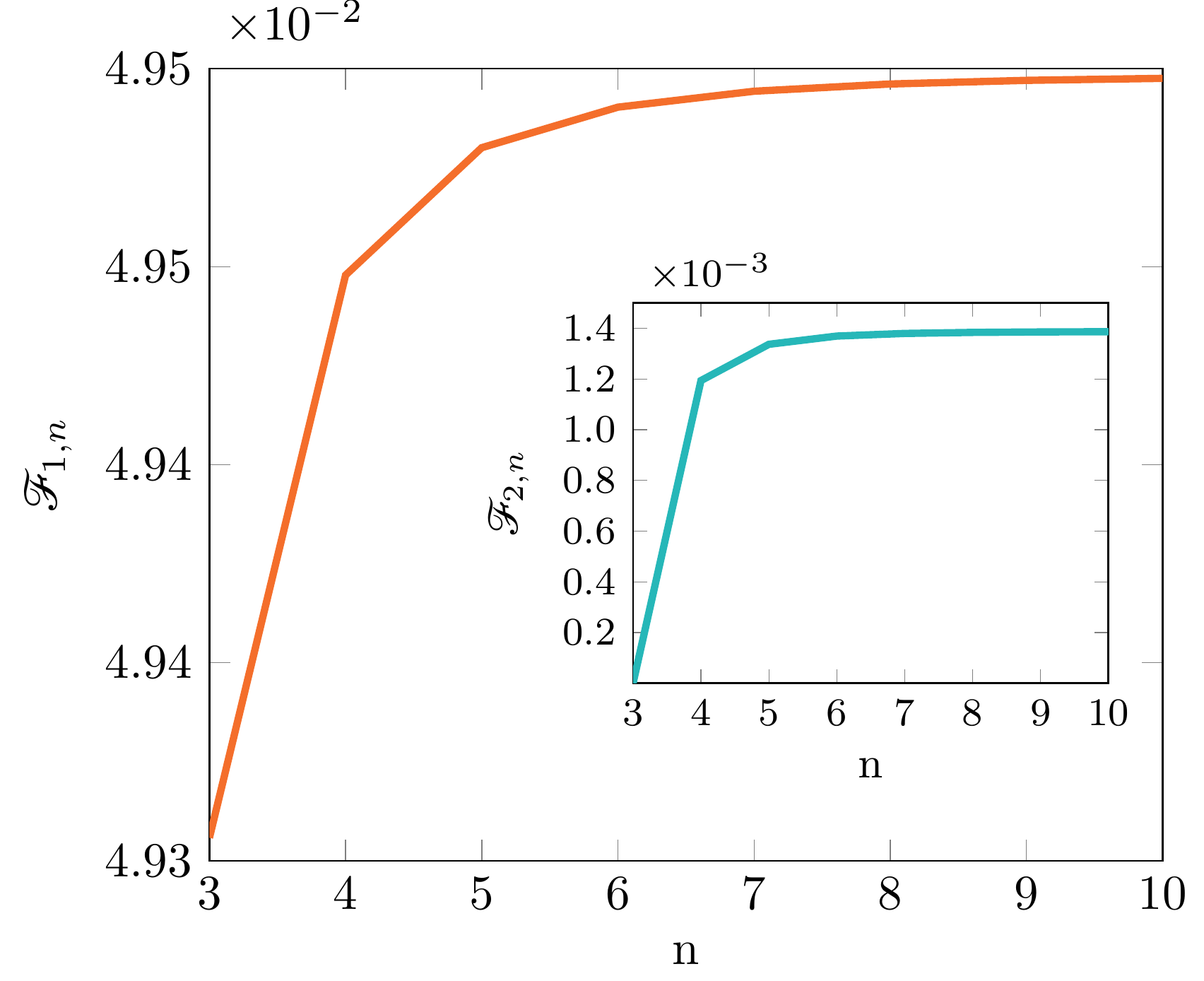}
	\caption{The Casimir force $\Fc_{s,\n}$ acting on the first and second (insert figure) graphene in the stack of $\n$ graphene. }
	\label{fig:fo2}
\end{figure}

The numerical evaluations of force acting on the graphene $s$ in the stack of $\n$ graphene are plotted in Figs.\,\ref{fig:fo1}, \ref{fig:fo2}. We evaluate the function $\Fc_{s,\n} = F_{s,\n}/F_{Cas}$, where $F_{Cas} = - \pi^2/240 d^4$ is the Casimir force for two ideal planes. 
From Fig.\,\ref{fig:fo1}, we observe that the value of the force falls very quickly, beginning at the first graphene sheet. The force acting on the second graphene is in $35$ time is smaller than the force acting on the first one.  Fig.\,\ref{fig:fo2} shows that the force very quickly becomes the force for infinite number of graphene -- already the stack of $10$ graphene gives the same force as for an infinite stack.

\section{Conclusion}

We considered here the Casimir energy for a stack of conductive planes and a force acting on the plane on this stack. To calculate the Casimir force, the scattering theory is used. Starting from the scattering on the single plane, we found the scattering matrix for the set of planes and found recurrent relation \eqref{eq:RecRelGen} for the Casimir energy \eqref{eq:CasGen}. In the particular case of a stack of graphene with equal interplane distance, the recurrent relation \eqref{eq:Dcn} may be solved in the manifest form \eqref{eq:DcnS}. The reflection Fresnel matrix may be diagonalized with eigenvalues corresponding to the TE and TM electromagnetic modes. The resulting expression for the Casimir energy is the sum of contributions of these modes \eqref{eq:EnGen}. The function $D_\n$ has the same form \eqref{eq:Dnx} which was found in Ref.\,\cite{Khusnutdinov:2015:cefasocp} for the constant conductivity case. We have to use, only the corresponding reflection coefficients for graphene \eqref{eq:FresnelGraphene}. 

For small interplane separations $md \ll 1$, the Casimir energy is divergent as $1/d^3$ (without mass gap parameter $m$). In the opposite limit of large distances, the energy falls as $1/(m^2 d^5)$ in accordance to Ref.\,\cite{Bordag:2009:CibapcagdbtDm}. The binding energy of infinite stack graphene with interplane separation of graphite, $d_c$, is $E_{ib} = 9.9 $ meV/atom, which is by $10$\% smaller than for constant conductivity case considered in Ref.\,\cite{Khusnutdinov:2015:cefasocp}. The obtained value of binding energy depends only weakly on the mass gap parameter because the case of graphite interplane separation corresponds to the case $md_c \ll1$ (see the insert in Fig.\,\ref{fig:enp}).

The Casimir force acting on the graphene $s$ in the stack of $\n$ graphene is a sum of two contributions \eqref{eq:Force}, where function $G_{s,\n}$ has the same form \eqref{eq:FnG} as for constant conductivity case found in Ref.\,\cite{Khusnutdinov:2015:cefasocp} with corresponding reflections coefficients for TE and TM modes. The force acting on graphene falls very quickly -- for the second plane it is $35$ times smaller than for the first graphene (see Fig.\,\ref{fig:fo1}).  The extremes graphenes squeezes the stack. The force acting on graphene in a stack is quick, $\n \sim 10$ (see Fig.\,\ref{fig:fo2}), becomes the same as for an infinite stack. 

The influence of non-zero temperature and chemical potential can be taken into account by employing the corresponding conductivity obtained in Ref.\,\cite{Fialkovsky:2011:FCefg,*Bordag:2016:ECefdg}. We will analyze this influence in forthcoming papers.

\begin{acknowledgments}
	NK was supported in part by the grants 2022/08771-5, 2021/10128-0 of S\~ao Paulo Research Foundation (FAPESP). 
\end{acknowledgments}

\appendix 
\section{Some relations}\label{Sec:app1}
\textbf{I)} First of all we prove the formula \eqref{eq:SK}. By using relations \eqref{eq:direct}, \eqref{eq:inverse} and \eqref{eq:Fresnel} one has
\begin{align}
	\det \Sb &= \frac{\det \Rb}{\det \ov{\Rb}'} = \frac{\det \Rb}{\det (\Rb' - \Rb \Rb^{-1} \Tb')^{-1}} \ann 
	&= \frac{\det\left[-\Kb^{-1}_{22}\Kb_{21}\right]}{\det\left[\Kb_{12}\Kb^{-1}_{22} + \left(\Kb_{11} - \Kb_{12} \Kb_{22}^{-1} \Kb_{21}\right)\Kb_{21}^{-1}\Kb_{22}\Kb_{22}^{-1}\right]^{-1}}\ann
	&= \frac{\det\left[-\Kb^{-1}_{22}\Kb_{21}\right]}{\det\left[\Kb_{11}\Kb_{21}^{-1}\right]^{-1}} = \frac{\det\left[-\Kb^{-1}_{22}\Kb_{21}\right]}{\det\left[\Kb_{21} \Kb_{11}^{-1}\right]} = \frac{\det \Kb_{11}}{\det \Kb_{22}}.
\end{align}

\textbf{II)} The different forms of the matrices $\Bb_i$. 

By direct solving the relation $\Cb^{-1}_i\Cb_i = \Ib$, we obtain
\begin{equation}
	\Cb^{-1}_i=
	\begin{pmatrix}
		\rb'_i\tb'^{-1}_i &  -\Ib\\
		\tb'^{-1}_i     &   0
	\end{pmatrix}.
\end{equation}

Therefore,  we can represent the matrix $\Bb_i$ in different forms
\begin{equation}\label{eq:Bdif}
	\Bb_i =
	\begin{pmatrix}
		- \ov{\tb}'^{-1}_i &  -\rb'_i \tb'^{-1}_i\\
		\tb'^{-1}_i \rb_i     &   -\tb'^{-1}_i
	\end{pmatrix}=
	\begin{pmatrix}
		- \ov{\tb}'^{-1}_i &  \ov{\tb}'^{-1}_i \ov{\rb}_i\\
		\ov{\rb}'_i \ov{\tb}'^{-1}_i     &   -\tb'^{-1}_i
	\end{pmatrix}=
	\begin{pmatrix}
		- \ov{\tb}'^{-1}_i &  \ov{\tb}'^{-1}_i \ov{\rb}'_i\\
		\tb'^{-1}_i \rb_i     &   -\tb'^{-1}_i
	\end{pmatrix},
\end{equation}
taking into account relations \eqref{eq:direct} and \eqref{eq:inverse}. Therefore, the matrices have the following structure
\begin{equation} \label{eq:Bsym}
	\Bb_i =
	\begin{pmatrix}
		\bb^i_{11} &  \bb^i_{12}\\
		\ov{\bb}^i_{12}     &  \ov{\bb}^i_{11}
	\end{pmatrix} =
	\begin{pmatrix}
		\ov{\bb}^i_{22} &  \ov{\bb}^i_{21}\\
		\bb^i_{21}     &  \bb^i_{22}
	\end{pmatrix} .
\end{equation}
With replacement \eqref{eq:replacement} the dependence of the position appears
\begin{equation}
	\Bb_i =
	\begin{pmatrix}
		\bb^i_{11} &  \bb^i_{12} e^{-2\ii k_z  z_i}\\
		\bb^i_{21} e^{2\ii k_z z_i}    &  \bb^i_{22}
	\end{pmatrix}.
\end{equation}
Some properties:
\begin{equation}
	\Bb_i^{-1} =
	\begin{pmatrix}
		-\tb^{-1}_i & \tb^{-1}_i \rb'_i \\
		-\rb_i \tb^{-1}_i & - \ov{\tb}^{-1}_i
	\end{pmatrix},\ \det \Bb_i = \frac{\det \tb_i}{\det \tb'_i}.
\end{equation}

Therefore, due to relations \eqref{eq:Bsym} the matrix $\Kb$ has the following structure
\begin{equation}\label{eq:KSym}
	\Kb =
	\begin{pmatrix}
		\Kb_{11} &  \Kb_{12}\\
		\ov{\Kb}_{12}     &  \ov{\Kb}_{11}
	\end{pmatrix} =
	\begin{pmatrix}
		\ov{\Kb}_{22} &  \ov{\Kb}_{21}\\
		\Kb_{21}     &  \Kb_{22}
	\end{pmatrix}.
\end{equation}

The matrices $\Bb_i$ maybe represented in the following forms
\begin{equation}
	\Bb_i =  \taub_i
	\rhob_i =
	\rhob'_i
	\taub_i,
\end{equation}
where
\begin{equation}
	\rhob_i =  	\begin{pmatrix}
		-\Ib & \ov{\rb}_i \\
		\rb_i & -\Ib
	\end{pmatrix}, \rhob'_i =
	\begin{pmatrix}
		-\Ib & -\rb'_i \\
		-\ov{\rb}'_i & -\Ib
	\end{pmatrix}, \taub_i =
	\begin{pmatrix}
		\ov{\tb}'^{-1}_i & 0 \\
		0 & \tb'^{-1}_i
	\end{pmatrix}.
\end{equation}
It easy to see that
\begin{equation}
	\rhob_{i+1} \rhob'_i =
	\begin{pmatrix}
		\Ib - \ov{\rb}_{i+1} \ov{\rb}'_i & \rb'_i - \ov{\rb}_{i+1}	\\
		\ov{\rb}'_{i} -\rb_{i+1} & \Ib - \rb_{i+1} \rb'_i
	\end{pmatrix}.
\end{equation}

\section{Different number of planes}\label{Sec:app2}
\subsection{The case without planes, $\n =0$}

Without planes there is no scattering and $\Kb_0 = \Ib_4$ is diagonal matrix $4\times 4$. We define $\Dc_0 = \Ib$.

\subsection{Single plane, $\n = 1$}

The matrix $\Kb_1 = \Bb_1$ has the following form (see Eq. \eqref{eq:Bdif})
\begin{equation}
	\Kb_1 =
	\begin{pmatrix}
		- \ov{\tb}'^{-1}_1 &  -\rb'_1 \tb'^{-1}_1\\
		\tb'^{-1}_1\rb_1     &   -\tb'^{-1}_1
	\end{pmatrix}.
\end{equation}

Therefore, 
\begin{equation}
	\Dc_1 = -\Kb_1^{22} \tb'_1 = \Ib.
\end{equation}
This expression does not depend on the position of plane and gives no contribution to the Casimir energy. 

\subsection{Two planes, $\n = 2$}

In this case
\begin{equation}
	\Kb_2= \Bb_2 \Bb_1 = \taub_2
	\begin{pmatrix}
		\Ib - \ov{\rb}_{2} \ov{\rb}'_1 & \rb'_1 - \ov{\rb}_{2}	\\
		\ov{\rb}'_{1} -\rb_{2} & \Ib - \rb_{2} \rb'_1
	\end{pmatrix} \taub_1.
\end{equation}
Therefore,
\begin{equation}
	\Dc_2 = \Ib - \rb_{2} \rb'_1.
\end{equation}

\subsection{Three planes, $\n =3$}

For three planes one has
\begin{equation}
	\Kb_3 =   \Bb_3 \Bb_2\Bb_1 =
	-\taub_3\begin{pmatrix}
		\Ib - \ov{\rb}_3 \ov{\rb}'_2 & \rb'_2 - \ov{\rb}_3	\\
		\ov{\rb}'_2 -\rb_3 & \Ib - \rb_3 \rb'_2
	\end{pmatrix}
	\begin{pmatrix}
		\ov{\tb}'^{-1}_2 & \ov{\tb}'^{-1}_2\rb'_1 \\
		\tb'^{-1}_2\ov{\rb}'_1 & \tb'^{-1}_2
	\end{pmatrix}\taub_1. \label{eq:three}
\end{equation}
Then, by using the relations 
\begin{equation}
	\ov{\rb}' \ov{\tb}'^{-1} =  - \tb'^{-1}\rb,\ \ov{\tb}'^{-1}\ov{\rb} = - \rb' \tb'^{-1}, \label{eq:Relations}
\end{equation}
and Eqs. \eqref{eq:direct}, \eqref{eq:inverse} we obtain
\begin{equation}\label{eq:E3Gen}
	\Dc_3 = \Dc_2^{(32)}\tb'^{-1}_2\Dc_2^{(21)}\tb'_2  - \rb_3\tb_2\rb'_1\tb'_2,
\end{equation}
where $\Dc_2^{(ij)} = \Ib - \rb_i \rb'_j$ is for two planes $i$ and $j$.  In manifest form
\begin{equation}
	\Dc_3 =\Ib - \rb_3 \rb'_2   - \tb'^{-1}_2 \rb_2\rb'_1\tb'_2 + \rb_3 \rb'_2\tb'^{-1}_2 \rb_2\rb'_1\tb'_2 - \rb_3\tb_2\rb'_1 \tb'_2.
\end{equation}
By using replacement \eqref{eq:replacement} we observe that the second term $\sim e^{2\ii k_z d_{3,2}}$, the third term $\sim e^{2\ii k_z d_{2,1}}$, and the last two terms $\sim e^{2\ii k_z d_{3,1}}$, where $d_{i,k}$ is distance between planes $i$ and $k$.

A comment is in order. The first term in expression obtained \eqref{eq:E3Gen} gives directly the sum of energies of pars. Indeed,
\begin{equation}
	\ln \det (\Dc_2^{(32)}\tb'^{-1}_2\Dc_2^{(21)}\tb'_2) = \ln \det \Dc_2^{(32)} + \ln \det \Dc_2^{(21)}.
\end{equation}
The second term breaks the additivity of the Casimir energy. It contains refraction coefficients the first and last plane without middle plane.

\subsection{Four planes}

We have 
\begin{align}
	\Kb_4 &=   \taub_4
	\begin{pmatrix}
		\Ib - \ov{\rb}_4 \ov{\rb}'_3 & \rb'_3 - \ov{\rb}_4	\\
		\ov{\rb}'_3 -\rb_4 & \Ib - \rb_4 \rb'_3
	\end{pmatrix}
	\taub_3 \taub_2
	\begin{pmatrix}
		\Ib - \ov{\rb}_2 \ov{\rb}'_1 & \rb'_1 - \ov{\rb}_2	\\
		\ov{\rb}'_1 -\rb_2 & \Ib - \rb_2 \rb'_1
	\end{pmatrix}\taub_1.
\end{align}
Let us express now this expression to that without asterisk. We use relations \eqref{eq:Relations}. Then,
\begin{align}
	\Dc_4 &= \Dc_2^{(43)}\tb'^{-1}_3 \Dc_2^{(32)} \tb'^{-1}_2\Dc_2^{(21)}\tb'_2 \tb'_3 - \Dc_2^{(43)}\tb'^{-1}_3\rb_3\tb_2\rb'_1\tb'_2 \tb'_3 \ann
	&-  \rb_4 \tb_3 \rb'_2 \tb'^{-1}_2 \Dc_2^{(21)}\tb'_2 \tb'_3 - \rb_4 \tb_3 \tb_2\rb'_1\tb'_2 \tb'_3.
\end{align}
The first term describes additive part of the Casimir energy and rest terms give non-additivity of the energy.

%

\end{document}